\documentclass[12pt,showpacs,preprint,eqsecnum,amsmath,amssymb]{revtex4}
\usepackage{graphicx}
\usepackage{amsmath}
\usepackage{amssymb}
\begin{document}
\title{FANO RESONANCES IN STUBBED QUANTUM \\
WAVEGUIDES WITH IMPURITIES}
\author{Giorgio Cattapan}
\affiliation{Dipartimento di Fisica ``G. Galilei'', Universit\`a di Padova,
Via F. Marzolo 8, I-35131 Padova, Italy \\ 
Istituto Nazionale di Fisica Nucleare, Sezione di Padova, 
Via F. Marzolo 8, I-35131 Padova, Italy}
\author{Paolo Lotti}
\affiliation{Istituto Nazionale di Fisica Nucleare, Sezione di Padova, 
Via F. Marzolo 8, I-35131 Padova, Italy \\
Dipartimento di Fisica ``G. Galilei'', Universit\`a di Padova,
Via F. Marzolo 8, I-35131 Padova, Italy} 
\begin{abstract}
We consider T--shaped, two--dimensional quantum waveguides containing 
attractive or repulsive impurities with a smooth, realistic shape,
and study how the resonance behavior of the total conductance depends upon 
the strength of the defect potential and the geometry of the device. 
The resonance parameters are determined locating the relevant $S$--matrix 
poles in the Riemann energy surface. The total scattering operator
is obtained from the $S$--matrices of the various constituent segments of the 
device through the $\bigstar$--product composition rule. This allows for a 
numerically stable evaluation of the scattering matrix and of the
resonance parameters.
\end{abstract}
\pacs{73.63.Nm \newblock{Quantum wires}, 73.23.Ad \newblock{Ballistic 
transport}, 72.10.Fk \newblock{Scattering by point defects, dislocations, 
surfaces, and other imperfections (including Kondo effect)}
\newline
\vfill
\hfill Preprint DFPD/07/TH11}
\maketitle
\section{\label{intro}Introduction}

	The discovery of conductance quantization in  
microcostrictions \cite{wee88,wha88} has prompted a  great deal of 
theoretical and  experimental activity on electron transport in such systems 
\cite{da95,fg99,lo99}. It has been soon realized that the presence of
an impurity in the waveguide can modify the shape of the conductance
dramatically, especially near the thresholds where
propagation modes are opened. These effects have been investigated
in many model calculations, for both point--like defects 
\cite{bag90,kcb96,blr00,blr00p} and more realistic, finite--range
defect shapes \cite{bag90,kis99,bam04,vap05}. The scattering of the
electron wave off the impurity produces resonance phenomena in the
conductance; in particular, for attractive impurities one has deep
transmission minima, which can be interpreted as due to the coupling
of a propagation mode in one subbband with a localized state formed
from another subband. This resonant suppression of ballistic
conductance is similar to the Fano resonances one observes in atomic
and nuclear physics, and has been the subject of detailed numerical 
\cite{kis99,bam04,vap05} and theoretical \cite{gl93,ns94} 
investigations. Needless to say, whereas theoretical arguments can
give insight into the general mechanisms underlying the formation
of resonances, only detailed numerical experiments can provide 
information about the dependence of resonance parameters 
upon the strength, shape, and position of the impurity potential inside
the quantum waveguide.

	The step--wise structure of conductance can be deeply influenced
also when the waveguide boundary is modified, with respect to an idealized, 
straight shape. In particular, the effects of a sidearm, or stub, attached
to the duct have been the subject of several investigations
\cite{mak91,wus91,shx96,sh97,der00,wv02,wrb05} since the pioneering papers of 
Sols {\em et al.} \cite{som89,som89l}, where a T--shaped waveguide was 
considered as a possible candidate for transistor action based on quantum 
interference. The conductance as a function of the Fermi energy exhibits
deep transmission minima, which can be again interpreted as reflection
resonances due to the excitation of quasi--bound states in the cavity
region \cite{der00}. A great deal of theoretical activity 
has been devoted to disentangle the physical factors influencing Fano
line shapes in these mesoscopic systems. Indeed, Fano resonances are 
interesting in themselves, because they are extremely sensitive to the 
details of the scattering process, and may be exploited to probe the degree 
of coherence in the scattering device \cite{getal00,cwb01,rl04,umm07}; at the 
same time, their strong dependence upon the energy of the scattered particle 
make them promising candidates for practical applications, such as the spin 
filtering action when spin degeneracy is removed by the application of an 
external magnetic field or by use of the Rashba effect \cite{sob03}.
At low level densities, and for systems coupled to very few channels,
quantum interference between quantum states plays a prominent role, and
a detailed quantum-mechanical study of the scattering device is required.
The dependence of the spectral properties of quantum dots attached to quantum 
waveguides upon the dot's dimensions has been the subject of detailed analysis
in recent years \cite{nr98,sbr06}.  

 	To the best of our knowledge, up to now an analysis of the
simultaneous effects of impurities and stubs in quantum interference
devices has never been accomplished. It is the purpose of this paper
to provide a first contribution, in order to fill this gap. In particular,
we have studied how the resonance features of the conductance depend
upon the strength of an embedded defect, and the size of the stub.
To this end, we have determined how the resonance poles of the scattering
operator for the device move on the multi--sheeted Riemann energy surface,
as the device parameters are varied. For stubbed waveguides with realistic 
defects, this requires an accurate numerical solution of the two--dimensional 
Schr\"odinger equation, in a large range of values for the stub's dimensions.  
This is made possible combining mode--matching techniques with an 
$S$--matrix approach. The device is regarded as a cascade 
of uniform waveguide sections, connected through interfaces or junctions, 
where a change in the transverse dimensions occurs, or the electron 
experiences a variation in the strength of the interaction. The scattering 
parameters of the device can be then determined starting from the scattering 
properties of the various building blocks through the $\bigstar$--product 
composition rule for the partial scattering operators \cite{da95,ki88,cm88}. 
Since forward and backward states are treated separately by the 
$S$--matrix, so that the propagating and less localized components
play a prominent role, numerical stability is guaranteed as some dimension of 
the system becomes large. At the same time, a varying number of basis 
functions can be accommodated in a natural way in the scattering matrix 
approach; this is particularly important in non--uniform waveguides, where one 
expects that a different number of wave--function components has to be 
considered as the transverse size of the device changes. Thus, the employment 
of the scattering operator, together with the $\bigstar$--product 
composition rule, opens the way to an accurate description of devices with 
a fairly large number of different constituent segments, as required
when dealing with impurities having a smooth profile along the 
propagation direction. They are in fact approximated by a cascade of 
several thin layers, in which the dependence upon the propagation coordinate is
neglected.     
 
	The paper is organized as follows. In Sect. \ref{forma} we review the 
scattering matrix approach to ballistic transport in quantum waveguides. In 
particular, we point out the compatibility between the $\bigstar$--product 
composition rule and the non--trivial block--wise structure of the partial 
$S$--matrices, when a different number of basis functions is chosen in 
different slices of the device. With reference to this point, we would like to 
point out that, whereas the numerical stability of the $S$--matrix has been 
widely recognized in the literature \cite{da95,ki88}, its flexibility with 
respect to the number of wave function components has been less stressed in 
the past, with the exception of Ref. \cite{wei91}. In Sect. \ref{numerics}
we give several examples of how the device parameters influence the pole
location in the energy--plane, and how this affects the resonant
behavior of the conductance. Our main conclusions are briefly summarized in 
Sect. \ref{conc}. 
           
\section{\label{forma}Scattering formalism for quantum transport in 
stubbed waveguides}

\begin{figure}
\centerline{\includegraphics[width=10 truecm,angle=0]{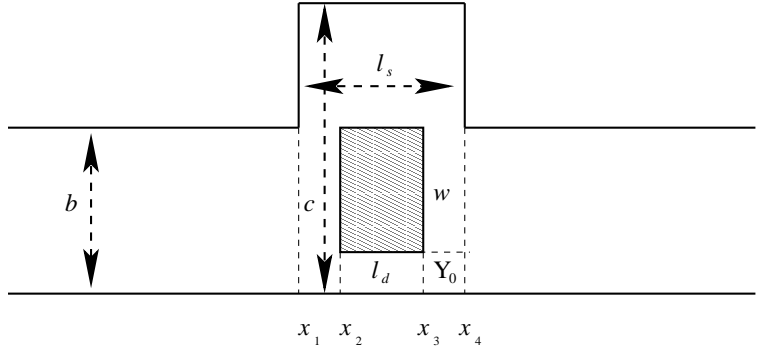}}
\caption{A stubbed quantum waveguide of width $b$ and infinite length, with
a stub of width $c$ and length $l_s$. The stub contains a defect with 
dimensions $w\times l_d$.}
\label{fig1}
\end{figure}

We consider the T--shaped device illustrated in Fig. \ref{fig1}. 
It consists of a uniform guide of indefinite length and width $b$, 
with a sidearm (stub) having width $c$ and length $l_s$. We shall 
assume simple hard wall boundary conditions, so that
the two--dimensional electron wave function has to vanish on the
boundary of the device. The stub may contain a region, where there
is a defect, or an external applied field. This region, to which we
shall simply refer as ``defect'', is exhibited by the shaded area in 
Fig. \ref{fig1}.  The electron's wave function $\Psi (x, y)$ is then the
solution of the two--dimensional Schr\"odinger equation for a given total 
energy $E$, the electron in the conduction band being endowed with an 
effective mass $m^\ast$. In the present paper we have chosen for 
$m^\ast$ the value $0.067 m_e$, which is appropriate for the 
${\rm Al}_x{\rm Ga}_{1-x}{\rm As}/{\rm GaAs}$ interface. The potential 
field due to the defect will be represented by an interaction term 
$V(x, y)$, whereas the effect of the confining potential is taken into 
account through the boundary conditions that $\Psi(x, y)$ vanishes along 
the edge of the straight duct and of the stub.
Because of the presence of the potential term $V(x, y)$, the full 
Schr\"odinger equation is not in general separable. As is well known 
\cite{lo99}, one can reduce it to an (in principle) infinite set of 
one--dimensional differential equations by expanding its solution into the 
transverse mode eigenfunctions $\phi_n^{(r)}(y)$ in the leads $(r=l)$ and in 
the stub $(r=s)$. Under the assumption of hard-wall boundary conditions, these 
basis functions are just the eigenfunctions of infinite square wells in the 
transverse direction, with widths $b$ and $c$ in the leads and in the stub, 
respectively.

   The whole device can be divided into five regions; the left and right ducts,
the two empty cavities in the sidearm, and the defect. The total wave function
in the different regions of the stub can be piecewise expanded in terms of the 
basis functions $\phi_n^{(s)}$. Once this expansion is inserted into the 
Schr\"odinger equation, one immediately finds that the expansion coefficients 
satisfy uncoupled differential equations, and can be written in terms of 
forward and backward propagating waves in the regions of the stub where the 
potential $V(x, y)$ is negligible; the corresponding wave numbers $k^{(s)}_n$ 
are related to the total energy $E$ and to the transverse eigenenergies by
\begin{equation}
k^{(s)}_n = \sqrt{\frac{2m^\ast}{\hbar^2}E-\left(\frac{n\pi}{c}\right)^2}
\equiv \sqrt{\frac{2m^\ast(E-\epsilon_n^{(s)})}{\hbar^2}}~.
\label{kappas}
\end{equation}
These relations essentially express the conservation of energy; once the 
$n$--th transverse eigenmode has been excited, only the energy 
$(\hbar^2/2m^\ast){k^{(s)}_n}^2$ is left for the electron propagation along 
the $x$--direction. If $E < \epsilon^{(s)}_n$, the wave number has to be taken
as purely imaginary, {\em i.e.}, $k^{(s)}_n \equiv i \kappa^{(s)}_n$, with
\begin{equation}
\kappa^{(s)}_n = \sqrt{\left(\frac{n\pi}{c}\right)^2 - 
\frac{2m^\ast}{\hbar^2} E}~~.
\label{chis}
\end{equation}
The associated propagating waves become then real, exponentially decaying or 
growing wave functions, and one has an evanescent mode, or closed channel.  

In presence of the defect a somewhat more involved procedure
is required, since one gets coupled equations for the expansion coefficients.
These equations can be straightforwardly uncoupled, and the coefficients 
written in terms of forward and backward propagating waves, if one assumes 
for the potential $V(x, y)$ the simple form
\begin{equation}
V(x, y) = V_0f(y)\Theta(x-x_2)\Theta(x_3-x)~~~,
\label{potsqua}
\end{equation}
where $\Theta(x-x_i)$ is the Heaviside step function. In such a case one 
arrives at the required result diagonalizing the  Hamiltonian 
$\left[- \frac{\hbar^2}{2m^\ast}\frac{d^2}{d y^2} + V_0f(y)\right]$
for the transverse motion, in the model space spanned by the basis functions 
$\phi^{(s)}_n(y)$  \cite{bag90}. In the potential region of the stub the 
expansion coefficients can then be written as linear superpositions
of plane--wave components $e^{\pm i k_n^{(d)}(x-x_2)}$, with wave vectors  
\begin{equation}
k^{(d)}_n = \frac{\sqrt{2m^\ast\left(E - {\cal E}_n\right)}}{\hbar}~~,
\label{newkappa}
\end{equation}
the quantities ${\cal E}_n$ being the eigenvalues of the transverse
Hamiltonian. Here also, if $E < {\cal E}_n$, the wave number is purely 
imaginary, in completely analogy to Eq. (\ref{chis}), and 
one has evanescent, non propagating modes in presence of the defect 
interaction.  
 
	Similar considerations can be repeated for the wave function
in the leads, the only difference being that now the transverse
eigenmodes $\phi_n^{(l)}$ have to be used. The wave vectors in the 
leads are given by expressions quite similar to Eqs. (\ref{kappas}) and
(\ref{chis}) for open and closed channels, respectively, with the
stub transverse eigenenergies $\epsilon^{(s)}_n$ replaced by the
lead corresponding quantities $\epsilon^{(l)}_n \equiv 
\displaystyle{\frac{\hbar^2}{2m^\ast}\left(\frac{n\pi}{b}\right)^2}$. 
For the time being, we shall not specify the actual boundary conditions the 
wave function has to satisfy asymptotically, and we let the presence of 
incoming and outgoing waves from both the left and from the right in the ducts.
We just limit ourselves to observe that, in correspondence to closed
channels, one must exclude incoming waves from the left or from the
right, in order to avoid divergent components in the wave function. 

	The unknown coefficients appearing in the expansion of the wave 
function in the various regions can be related to one another by matching the 
wave function and its first derivative at the various interfaces. The linear 
transformation yielding the coefficients on one side of an  
interface in terms of those on the other side represents what is 
usually referred to as the transfer matrix \cite{lo99}. It is well--known 
that the transfer matrix suffers from several limitations 
\cite{ki88,xu94}. First of all, its generalization to the case where 
there is a different number $N_L$ of basis functions on the left of the 
matching surface with respect to the right is not straightforward. This is 
particularly relevant at the lead/stub or stub/lead interfaces, where a 
different number of open and closed channels may occur in the cavity with 
respect to the lead. A second problem arises in presence of evanescent modes, 
whose occurrence makes the inversion of the transfer matrix numerically 
troublesome when one evaluates the transmission coefficients as the dimensions 
of the device get large. The above shortcomings can be all overcome resorting 
to the scattering matrix \cite{da95,fg99}. In such a case, one expresses the 
outgoing amplitudes in terms of the incoming ones, to get  
\begin{equation}
\left(
\begin{matrix}
\mathbf{S}_{11}~ & ~\mathbf{S}_{12} \\
\noalign{\vspace{5 pt}}
\mathbf{S}_{21}~ & ~\mathbf{S}_{22}
\end{matrix}
\right)
=
\left(
\begin{matrix}
\mathbf{\Gamma}^{(L)}_+\mathbf{S}^{(I)}_{11}\mathbf{\Gamma}^{(L)}_+~ & 
~\mathbf{\Gamma}^{(L)}_+\mathbf{S}^{(I)}_{12} \\
\noalign{\vspace{5 pt}}
\mathbf{S}^{(I)}_{21}\mathbf{\Gamma}^{(L)}_+~                        &
~\mathbf{S}^{(I)}_{22}
\end{matrix}
\right)~~,
\label{smat}
\end{equation}
where the sub-matrices $\mathbf{S}^{(I)}_{ij}$ are given by
\begin{subequations}
\begin{eqnarray}
\mathbf{S}^{(I)}_{11} = &
\mathbf{F}
\left(
\mathbf{A}_{21}-\mathbf{B}_{21}\mathbf{B}^{-1}_{11}\mathbf{A}_{11}
\right)~~,
\label{reds11} \\
\mathbf{S}^{(I)}_{12}  = &
\mathbf{F}
\left(
\mathbf{B}_{22}+\mathbf{B}_{21}\mathbf{B}^{-1}_{11}\mathbf{B}_{12}
\right)~~,
\label{reds12} \\
\mathbf{S}^{(I)}_{21} = &
\overline{\mathbf{F}}
\left(
\mathbf{A}_{11}+\mathbf{A}_{12}\mathbf{A}^{-1}_{22}\mathbf{A}_{21}
\right)~~,
\label{reds21} \\
\mathbf{S}^{(I)}_{22} = &
\overline{\mathbf{F}}
\left(
- \mathbf{B}_{12}+\mathbf{A}_{12}\mathbf{A}^{-1}_{22}\mathbf{B}_{22}
\right)~,
\label{reds22}
\end{eqnarray}
\end{subequations}
with
\begin{subequations}
\begin{eqnarray}
\mathbf{F} \equiv & 
\left(
\mathbf{A}_{22}+\mathbf{B}_{21}\mathbf{B}^{-1}_{11}\mathbf{A}_{12}
\right)^{-1}\phantom{~~.}\\
\overline{\mathbf{F}} \equiv &
\left(
\mathbf{B}_{11}+\mathbf{A}_{12}\mathbf{A}^{-1}_{22}\mathbf{B}_{21}
\right)^{-1}~~.
\end{eqnarray}
\end{subequations}

   The actual structure of the matrices $\mathbf{A}_{ij}$ and $\mathbf{B}_{ij}$
depends upon the matching conditions one is considering. Whatsoever these
conditions may be, one does not need to assume the same number of 
basis functions on the two sides of the matching surface, since the
building blocks of the total $S$-matrix are not required
to be square. As a matter of fact, if $N_L$ wave function components are 
retained on the left, whereas $N_R$ basis functions are considered on the 
right, the whole sub-matrix $\mathbf{S}_{11}$ has 
dimensions $N_L\times N_L$, and $\mathbf{S}_{22}$ is a square $N_R \times N_R$ 
array, whereas $\mathbf{S}_{12}$ and $\mathbf{S}_{21}$ have dimensions 
$N_L\times N_R$ and $N_R\times N_L$, respectively. Overall, the total 
$S$-matrix expresses the $N_L+N_R$ outgoing amplitudes in terms of the 
$N_L+N_R$ coefficients of the incoming waves. As for the diagonal matrix 
$\mathbf{\Gamma}^{(L)}_{+}$, it takes into account the propagation of
the wave function in the slice on the left of the matching surface, and
is defined according to
\begin{equation}
\{\mathbf{\Gamma}^{(L)}_{+}\}_{nm} = 
e^{+i k^{(L)}_n \Delta x} \delta_{nm}~~,
\label{defgam}
\end{equation}
where $\Delta x$ is the width of the slice, and $k^{(L)}_n$ are the 
corresponding propagation wave numbers.  It is worth to observe that 
in the scattering matrix only the phase factors 
$\displaystyle{e^{+ i k^{(L)}_n \Delta x}}$ appear. In correspondence to an 
evanescent mode, one actually has the exponentially decaying factors 
$\displaystyle{e^{- \kappa^{(L)}_n  \Delta x}}$, which are no longer mixed up 
with the growing functions $\displaystyle{e^{\kappa^{(L)}_n \Delta x}}$. As 
the numerical calculations confirm, these features make the whole formalism 
numerically stable even when some characteristic length of the system, and in 
particular the stub's width $c$, gets large.

    Finally, we give the actual structure of the matrices $\mathbf{A}_{ij}$ 
and $\mathbf{B}_{ij}$ at the lead/stub and cavity/defect interfaces.
For the former one has 
$\mathbf{A}_{11} = \mathbf{A}_{12} = \mathbf{J}$, $\mathbf{B}_{11} = 
\mathbf{B}_{12} = \mathbf{1}$, and $\mathbf{A}_{21} = \mathbf{A}_{22} = 
\mathbf{K}^{(l)}$, $\mathbf{B}_{21} = \mathbf{B}_{22} = \tilde{\mathbf{J}}
\mathbf{K}^{(s)}$. The $N_R\times N_L$ matrix $\mathbf{J}$ contains the 
overlap integrals among the different basis functions on the left and on the 
right of the discontinuity,
\begin{equation}
\mathbf{J}_{mn} = \int_0^b{\phi_n^{(l)}(y) \phi_m^{(s)}(y)dy}~~,
\label{overlap}
\end{equation}
$\tilde{\mathbf{J}}$ is the transpose of $\mathbf{J}$, while 
$\mathbf{K}^{(l)}$ and $\mathbf{K}^{(s)}$ are $N_L\times N_L$ and 
$N_R\times N_R$ diagonal matrices, whose diagonal elements represent the wave 
vectors $k_n^{(l)}$ and $k_n^{(s)}$, in the lead and in the stub, respectively.
For the cavity/defect interface, on the other hand, one has
$\mathbf{A}_{11} = \mathbf{A}_{12} = \mathbf{1}$, 
$\mathbf{B}_{11} = \mathbf{B}_{12} = \mathbf{Z}$,  
and $\mathbf{A}_{21} = \mathbf{A}_{22} = \mathbf{K}^{(s)}$,
$\mathbf{B}_{21} = \mathbf{B}_{22} = \mathbf{Z}\mathbf{K}^{(d)}$,
where $\mathbf{Z}$ is the matrix diagonalizing the transverse Hamiltonian
in the interaction region, and the diagonal matrix $\mathbf{K}^{(d)}$ 
contains the wave vectors $k^{(d)}_n$ defined by Eq. (\ref{newkappa}). For 
the sake of simplicity, we have assumed an equal number of basis functions on 
the left and on the right of the cavity/defect boundary; as a matter of fact, 
the potential $V(x, y)$ in general is not so strong, as to perturb the 
transverse eigenvalues in an essential way, and good convergence can be 
achieved with an equal number of wave function components outside and inside 
the defect.

	The unperturbed propagation along the slice preceding
the interface can be put in better evidence by re-writing Eq.
(\ref{smat}) as a suitable combination of elementary scattering
operators. This can be achieved through the $\bigstar$--product
of $S$--matrices \cite{da95,ki88,cm88}, which expresses the overall 
scattering matrix $\mathbf{S}$ in terms of the partial scattering 
matrices $\mathbf{S}^{(a)}$ and $\mathbf{S}^{(b)}$ as
\begin{equation}
\mathbf{S} = 
\left(
\begin{matrix}
\mathbf{S}_{11}~ & ~\mathbf{S}_{12} \\
\noalign{\vspace{5 pt}}
\mathbf{S}_{21}~ & ~\mathbf{S}_{22}
\end{matrix}
\right) =
\mathbf{S}^{(a)} \bigstar \mathbf{S}^{(b)}~~,
\label{starp}
\end{equation}       
where
\begin{subequations}
\begin{eqnarray}
\mathbf{S}_{11}  = &\mathbf{S}^{(a)}_{11} + \mathbf{S}^{(a)}_{12}
\mathbf{S}^{(b)}_{11}\left(\mathbf{1} - \mathbf{S}^{(a)}_{22}
\mathbf{S}^{(b)}_{11}\right)^{-1}\mathbf{S}^{(a)}_{21}~~, 
\label{str11} \\
\mathbf{S}_{12} = & \phantom{\mathbf{S}^{(a)}_{11} + \mathbf{S}^{(a)}_{12}}
\mathbf{S}^{(a)}_{12} \left(\mathbf{1} - \mathbf{S}^{(b)}_{11}
\mathbf{S}^{(a)}_{22}\right)^{-1}\mathbf{S}^{(b)}_{12}~~, 
\label{str12} \\
\mathbf{S}_{21}  = & \phantom{\mathbf{S}^{(a)}_{11} + \mathbf{S}^{(a)}_{12}}
\mathbf{S}^{(b)}_{21} \left(\mathbf{1} - \mathbf{S}^{(a)}_{22}
\mathbf{S}^{(b)}_{11}\right)^{-1}\mathbf{S}^{(a)}_{21}~~, 
\label{str21} \\
\mathbf{S}_{22}  = & \mathbf{S}^{(b)}_{22} + \mathbf{S}^{(b)}_{21}
\mathbf{S}^{(a)}_{22}\left(\mathbf{1} - \mathbf{S}^{(b)}_{11}
\mathbf{S}^{(a)}_{22}\right)^{-1}\mathbf{S}^{(b)}_{12}~~. 
\label{str22}
\end{eqnarray}
\end{subequations}

	It is worth to observe that this composition rule 
does not require the sub-matrices $\mathbf{S}^{(a)}_{ij}$ and  
$\mathbf{S}^{(b)}_{ij}$ to be square matrices. In particular, the 
$S$--matrix appearing in Eq. (\ref{smat}) can be factorized as follows 
\begin{equation}
\mathbf{S} =
\left(
\begin{matrix}
\mathbf{0}              & \mathbf{\Gamma}^{(L)}_+ \\
\noalign{\vspace{5 pt}}
\mathbf{\Gamma}^{(L)}_+ & \mathbf{0} 
\end{matrix}
\right)\bigstar
\left(
\begin{matrix}
\mathbf{S}^{(I)}_{11}~ & ~\mathbf{S}^{(I)}_{12} \\
\noalign{\vspace{5 pt}} 
\mathbf{S}^{(I)}_{21}~ & ~\mathbf{S}^{(I)}_{22} \\  
\end{matrix}
\right)~~ \equiv \mathbf{\Gamma}^{(L)}\bigstar\mathbf{S}^{(I)}
\label{fact}
\end{equation}
where the block--wise anti--diagonal factor $\mathbf{\Gamma}^{(L)}$ 
takes into account the free propagation through the slice on the left 
of the interface, and the reduced $S$--matrix $\mathbf{S}^{(I)}$
exhibits the effects on the wave function of the matching at the
interface between the two considered regions of the device. The former 
is a $2N_L \times 2N_L$ matrix, whereas the latter has dimensions
$(N_L+N_R)\times (N_L+N_R)$, as detailed above.

	Eqs. (\ref{starp}) and (\ref{fact}) have been systematically 
employed to evaluate the total $S$--matrix, starting from the 
partial $S$--matrices at the lead/stub and cavity/defect boundaries, 
as well as from the scattering operators for the free propagation of the 
electron along the various slices of the device. As
noted above, while there are physical reasons to choose a different
number $N_l$ of transverse basis functions in the lead with respect
to the sidearm, one can choose the same number $N_s$ of components 
in the empty slices of the stub and inside the defect. Thus, one
has an $(N_l + N_s)\times (N_l + N_s)$ partial $S$--matrix 
$\mathbf{S}{(l|s)}$ at the lead/stub boundary, and a $(2N_s\times 2N_s)$
partial $S$--matrix $\mathbf{S}{(c|d)}$ for the cavity/defect propagation.
One can verify by inspection that, when the two operators are combined 
through the composition rule (\ref{starp}), one gets again a
matrix with dimensions $(N_l + N_s)\times (N_l + N_s)$. Things are
a bit different when one considers the stub/lead interface on the
right of the device. We shall not enter here into the detailed
structure of the stub/lead interface $S$--matrix. We limit ourselves to note 
that, once the lead/stub $S$--matrix ${\mathbf{S}(l|s)}^{(I)}$  
has been evaluated, the corresponding $S$--matrix ${\mathbf{S}(s|l)}^{(I)}$ 
on the right can be obtained without further computational effort, through
the correspondence rule  
\begin{equation}
\left(
\begin{matrix}
{\mathbf{S}(s|l)}^{(I)}_{11}~ & ~{\mathbf{S}(s|l)}^{(I)}_{12} \\
\noalign{\vspace{10 pt}}
{\mathbf{S}(s|l)}^{(I)}_{21}~ & ~{\mathbf{S}(s|l)}^{(I)}_{22} 
\end{matrix}
\right) = 
\left(
\begin{matrix}
{\mathbf{S}(l|s)}^{(I)}_{22}~ & ~{\mathbf{S}(l|s)}^{(I)}_{21} \\
\noalign{\vspace{10 pt}}
{\mathbf{S}(l|s)}^{(I)}_{12}~ & ~{\mathbf{S}(l|s)}^{(I)}_{11} 
\end{matrix}
\right)~~.
\label{corr}
\end{equation}
A similar result holds for the $S$--matrices at the cavity/def\-ect and
defect/cavity interfaces. As a consequence of Eq. (\ref{corr}),  
once the $(N_l+N_s)\times(N_l+N_s)$ partial 
$S$--matrix describing the propagation inside the device is combined 
with the $(N_s+N_l)\times(N_s+N_l)$ $S$--matrix $\mathbf{S}(s|l)$ 
through the $\bigstar$-product composition rule (\ref{starp}), one 
gets an overall $2 N_l\times 2 N_l$ scattering operator $\mathbf{S}^{(T)}$, 
which effectively takes into account the flux percolation through open
and closed channels in the sidearm.

  Potentials having a non--trivial $x$ dependence can be accommodated in our 
approach through a {\it slicing} technique \cite{vap05,shx96}. This is 
particularly easy to do for interactions which are factorisable in their $x$ 
and $y$ dependence, such as the double Gaussian
\begin{equation}
V(x, y) = V_0 e^{- \beta^2 \left(x - x_c\right)^2 - 
\alpha^2\left(y - y_c\right)^2} .
\label{doubga}
\end{equation}
If $x_{0}$ and $x_{F}$ are two points, where the value of the potential is 
negligible, we divide the interval $[x_{0},\; x_{F}]$ into $N$ 
sub--intervals of equal width $\Delta x = \left(x_{F} - x_{0}\right)/N$,
and mimic the smooth potential $V(x, y)$ by a sequence of pseu\-do--defects, 
each of width $\Delta x$ and strength 
$$V_0(i) \equiv 
V_0 e^{- \beta^2 \left(x_i - x_c\right)^2}\qquad (i=0,\ldots,N)~,$$ 
with $x_i = x_0 + i \Delta x$. The conductance of the waveguide in presence of 
the smooth potential (\ref{doubga}) is then evaluated for increasing $N$, 
until convergence of the results is obtained.

	Let us finally fix the actual boundary conditions under
which we are solving the scattering problem. We shall assume that
an electron impinges on the device moving from the left, with no
flux coming from the right. If one assumes 
an incoming wave of unit flux in a given propagation mode, then
the matrix elements $\left(\mathbf{S}^{(T)}_{11}\right)_{nm}$ represent
the reflection coefficients toward the left from the initial channel
$m$ into the final propagation mode $n$, while the quantities 
$\left(\mathbf{S}^{(T)}_{21}\right)_{nm}$ give the transmission 
amplitudes to the right from mode $m$ into mode $n$. Finally, for
a two--probe device the total conductance $G$ can be evaluated through
the B\"uttiker formula \cite{da95,fg99,bi85}
\begin{equation}
G =  \frac{2e^2}{h}\sum_{m,n}\frac{k^{(l)}_n}{k^{(l)}_m}
\left |\left(\mathbf{S}^{(T)}_{21}\right)_{nm} \right |^2~~,
\label{cond}
\end{equation}
where the summation is restricted to the open channels in the leads.

\section{\label{numerics}Selected numerical examples}

Our aim is to study the dependence of resonance parameters in a stubbed 
quantum wave guide upon the strength of an embedded defect and the geometry 
of the device, with particular reference to the stub height. We shall do this 
by locating the relevant S-matrix poles in the multi-sheeted energy plane, 
with cuts starting along the real axis from the various scattering thresholds.
From now on, all lengths are measured in units of the waveguide width $b$, 
and energies in terms of the waveguide's fundamental mode $\epsilon_1^{(l)}=
\displaystyle{\frac{\hbar^2}{2m^{\ast}}\left(\frac{\pi}{b}\right)^2}$,
so that the various thresholds are $\tilde E_n =  1,\,4,\,9,\,\ldots$.
Sheets will be specified giving the sign of the imaginary part of the
lead's momenta in each channel \cite{lw67}. Thus, referring to a four--channel
situation, the symbol $(++++)$ denotes the physical sheet, where the 
imaginary parts of all the channel momenta $k^{(l)}_i$ are positive,
whereas on sheet $(-+++)$ one has ${\rm Im} k^{(l)}_1 < 0$, and
${\rm Im} k^{(l)}_i > 0$ for the other three channels. Our calculations
refer to a duct with a double Gaussian defect described
by Eq. (\ref{doubga}). In all cases we have chosen 
$\tilde x_0 \equiv x_0/b = 0$, and $\tilde x_f = 1$, so that the region 
where the potential acts has a length $\tilde l_d = 1$. Moreover, we have 
set $\tilde w = 0.3$; for a wave guide of width $b=100$ \AA\ , the interaction 
region is therefore $30$ \AA\ wide and $100$ \AA\ long. The defect is placed 
at $\tilde Y_0 = 0.1$ from the lower edge of the guide. The decay constants 
along the transverse and the propagation direction have been fixed at 
$\tilde \alpha \equiv b \alpha= 15$ and $\tilde \beta = 10$, respectively,
so as to ensure that the potential is negligible outside the interaction
region. We have checked the convergence of the slicing method, and found 
that quite stable results are obtained with $N=10\div 15$ sub--intervals.

	For a waveguide attached to a stub of height $\tilde c \equiv c/b$ 
and length $\tilde l_s \equiv l_s/b = 1$, one can expect that a larger 
number of basis functions is required in the stub than in the guide to 
achieve convergence, this difference increasing with increasing $\tilde c$. 
As a matter of fact, for a given electron's energy $\tilde E$ there are less 
open channels in the waveguide than in the stub, the thresholds in the two 
regions being $\tilde E^{(l)}_n = n^2$ and  $\tilde E^{(s)}_n = 
\tilde E^{(l)}_n/{\tilde c}^2$, respectively. It follows that, for a given 
energy $\tilde E$, the number of open channels in the stub scales as
$\tilde c \sqrt{\tilde E}$, whereas in the guide it scales as 
$\sqrt{\tilde E}$. Since a sensible calculation has to include at least all 
the propagating modes, a larger number of basis functions is required in the 
stubbed region. As for the evanescent modes, one may recall that a 
closed--channel component is the less effective the shorter its attenuation 
length $\tilde \xi_n \equiv 1/\left(b\kappa_n\right)$ \cite{shx96}. The 
attenuation lengths in the lead and in the cavity are given in terms of the 
total energy 
$\tilde E$ by
\begin{equation}
\tilde \xi_n^{(l)} = \frac{1}{\pi\sqrt{n^2-\tilde E}}\;,\;
\tilde \xi_n^{(s)} = 
\frac{1}{\pi\sqrt{\left(\frac{n}{\tilde c}\right)^2-\tilde E}}~~.
\label{atten}
\end{equation}
From Eqs. (\ref{atten}), one has that two evanescent modes $n$ and $m$, in 
the external region and in the stub, respectively, have the same attenuation 
length if the corresponding quantum numbers are related by $m = \tilde c n$. 
As a consequence, if $3\div 4$ components are required in the lead, 
$\sim 12$ closed channels have to be included  in the stub, to assure
convergence. Our numerical results are consistent with these estimates for
both the conductance and the pole locations in the complex energy plane.
We have obtained stable results with four components in the
lead, and $10 \div 12$ basis functions in the stub for $1 \leq \tilde c
\leq 4$. In these conditions, the position of the resonance poles in the 
$E$-plane can be guaranteed with an accuracy of the order $10^{-5}$. Only
the bound--state poles turned out to be more sensitive to the number
of employed basis functions in the interaction region, which was to be
expected. In this latter case, we employed four channels in the duct and
up to 20 modes in the internal region.  

\begin{figure}
\centerline{\includegraphics[width=8 truecm,angle=-90]{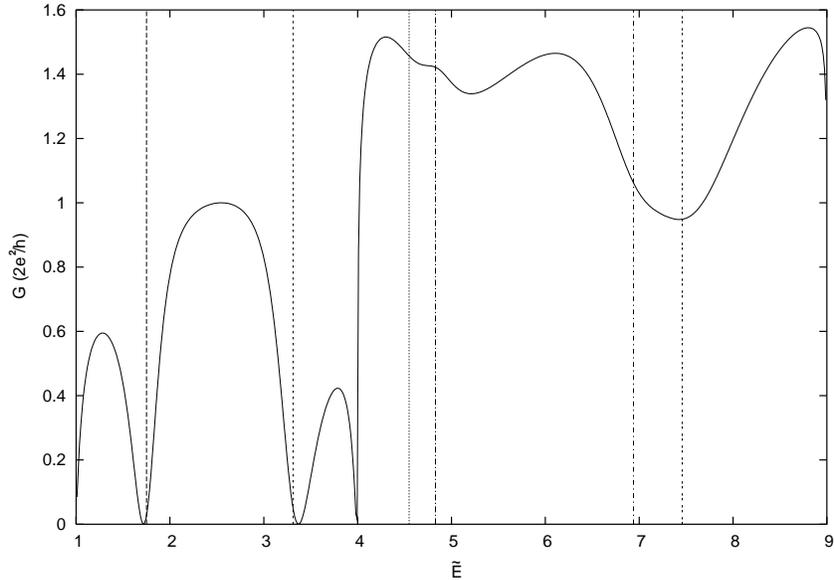}}
\caption{Conductance (in units $2e^2/h$) of a waveguide with a stub 
$\tilde c = 2$ high and a double--Gaussian defect of strength 
$\tilde V_\circ = 20$, in the energy region $1 \leq \tilde E \leq 9$.
The vertical lines show the value of the real part of the pole positions
in the complex energy plane. The poles lie in the $(-+++)$ and $(--++)$
sheet in correspondence to the first and second subband,respectively.}
\label{fig2}
\end{figure}

   The conductance of stubbed waveguides is characterized 
by a more or less complex pattern of transmission maxima and minima when 
considered as a function of the electron's energy. In Fig. \ref{fig2}
we give the conductance of a waveguide having $\tilde c =2$, with
a repulsive defect of strength $\tilde V_\circ = 20$, in the energy
region corresponding to the first two subbands $(1 \leq \tilde E \leq
9)$. 
One notes two transmission zeros in the first subband, and a non-trivial
oscillatory behavior with two large dips in the second one. We looked
for poles of the $S$--matrix in the unphysical sheets $(-+++)$ and
$(--++)$ and found two resonance poles at $\tilde E \simeq 1.75-0.19i$ and
$\tilde E \simeq 3.31-0.26i$ on the former sheet, and two pairs of neighboring
poles in the $(--++)$ sheet. The value of their real parts is exhibited by the
vertical dashed lines in Fig. \ref{fig2}. The two poles close to the first 
subband, together with the two transmission zeros, concur to produce the 
strong variations in the conductance one observes there. As for the poles 
on the $(--++)$ sheet, the first pair corresponds to the two almost 
overlapping  peaks one sees just above the lower edge of the second subband, 
whereas the other two poles give rise to the broad resonances at the top of 
this energy region. It is worth to note that the maxima of the conductance
are {\em not} aligned with the pole positions, which are very close to the
transmission zeros in the first subband. The conductance behavior in the
lowest subband is consistent with what one observes in simple, 
single--channel models for stubbed waveguides, where transmission resonances
arise because of the presence of zero--pole pairs in the complex energy
plane \cite{spl94}. In the higher subband, one clearly has a strong coupling   
between the occurring resonances.

\begin{figure}
\centerline{\includegraphics[width=8 truecm,angle=-90]{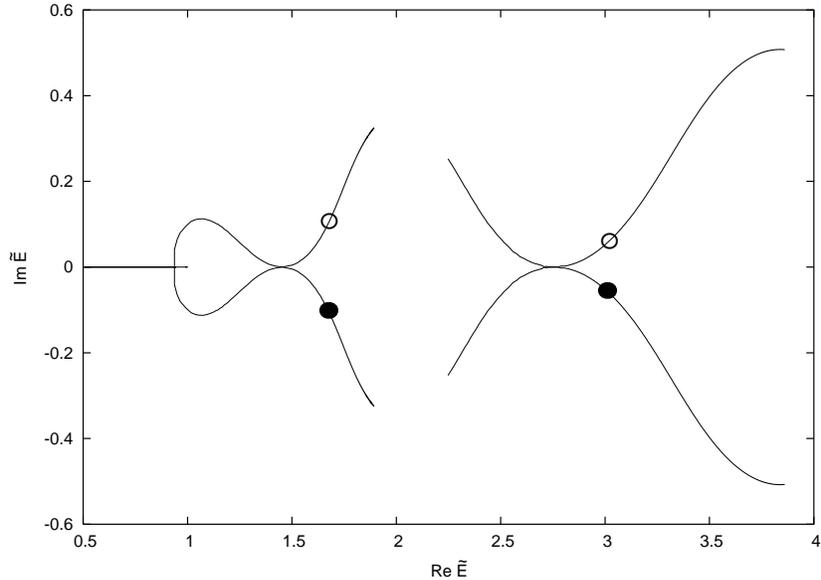}}
\caption{Trajectories of the $S$--matrix poles in the $(-+++)$ sheet  of
the complex energy plane for $-60 \leq \tilde V_\circ \leq 200$ and a
stub's height $\tilde c =2$. The full and empty dots mark the empty--stub
positions of the resonance poles and their complex conjugates, respectively.}
\label{fig3}
\end{figure}

  We have studied how the resonance poles move as the defect strength varies
with respect to the unperturbed situation, with the interaction switched
off. Limiting ourselves to the first subband, for the sake of simplicity, 
we give the outcome of our calculations in Fig. \ref{fig3}, for a stub
of height $\tilde c = 2$ and the strength varying in the interval
$-60 \leq \tilde V_\circ \leq 200$. For each resonance pole in the fourth 
quadrant of the energy plane, one observes a complex conjugate pole in the 
first quadrant. This is a confirmation of the correctness of our calculations.
Indeed, the multichannel $S$--matrix, when regarded as a function of the 
channel momenta $k_i$, has the basic property
$S(k_i) = S^\ast(-k^\ast_i)$ \cite{bk82}, 
which implies that to a pole located at $\tilde E = \tilde E^{(p)}$ there 
corresponds a pole at the complex conjugate energy 
$\tilde E = \tilde E^{(p)^\ast}$.
  As the defect strength becomes more and more repulsive, both resonance
poles in the fourth quadrant move upwards in energy, starting from the
empty-stub positions, marked by the full dots in Fig. \ref{fig3}. At the
same time they move away from the real axis; the presence of a repulsive
interaction displaces the resonance peaks at higher energy, as it was
to be expected. The broadening of the resonances with increasing 
defect strength is consistent with the calculations of Ref. \cite{cwb01},
where the effects of an impurity near the opening of a quantum dot
was considered, in an attempt to simulate the non-monotonic resonance
width found in a previous experiment \cite{getal00}. This effect can
be attributed to the deflection of more and more electrons into the
dot, as $\tilde V_\circ$ increases, so that the coupling between the
waveguide and the cavity gets stronger \cite{cwb01}. Stated in 
equivalent terms, one may say that the wave function trapping in the region 
where the waveguide and the duct match decreases at the defect 
becomes more repulsive. In the range of strengths we have explored $(0 \leq 
\tilde V_\circ \leq 200)$, on the other hand, we found no evidence of a 
non-monotonic behavior of the resonance width. For $\tilde V_\circ > 200$
the pole at higher energy moves above the upper edge of the first subband
$\tilde E_2$, and becomes a shadow pole \cite{et64}. As for the low--lying
pole, we have pushed the strength $\tilde V_\circ$ up to $1000$, 
finding no signal of a decreasing width. The reason for this difference
with respect to \cite{cwb01} can be traced back to the different role
played by backscattering in the two cases. In Ref. \cite{cwb01} the
impurity is near the opening of the dot, whereas in our case it is
embedded in the lower half of the waveguide. The pole trajectories
in the complex energy plane depend in a rather sensitive way upon
the position of the defect in the device, as we have found in 
previous model calculations \cite{clp06}. If the defect is displaced
closer to the dot, backscattering can have an increased role, and a
non-monotonic behavior of the resonance width may emerge. 

\begin{figure}
\centerline{\includegraphics[width=8 truecm,angle=-90]{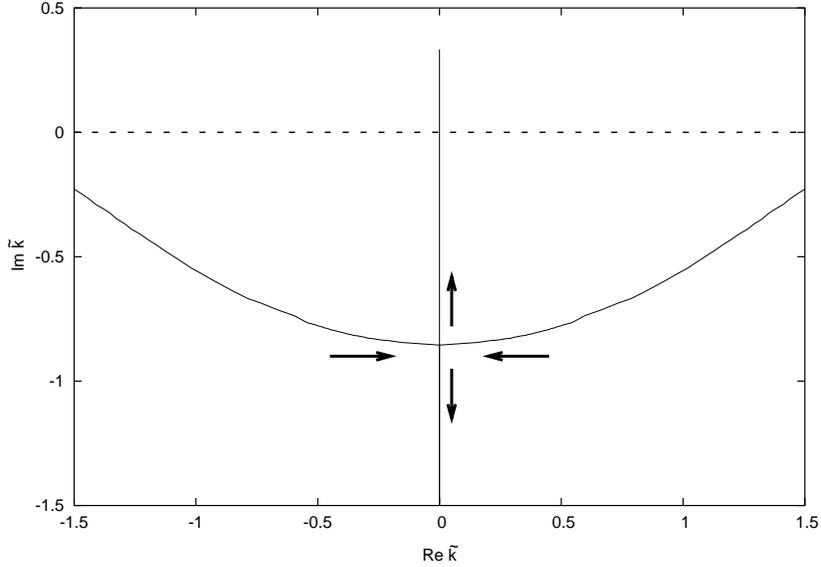}}
\caption{Motion of a pair of bound-state/anti-bound-state poles 
in the complex $k^{(l)}_1$--plane with varying strength in a stubbed
waveguide with $\tilde c =2$. The arrows display the movement direction as the
potential becomes more and more attractive.}
\label{fig4}
\end{figure}

For negative strengths, the two poles become closer and closer to the real 
axis; the first pole touches the real axis at $\tilde E \simeq 1.45$ for 
$\tilde V_\circ \simeq -15.6$, while the higher pole trajectory touches it at 
$\tilde E \simeq 2.76$ when $\tilde V_\circ \simeq -8.4$. Within the accuracy 
of the numerical calculations, we found that each pole lies just on the real 
axis in these conditions, at energies in the continuum of the first 
transmission band. This is an indication that for $\tilde V_\circ \simeq -8.4$
and $\tilde V_\circ \simeq -15.6$ one has a bound state embedded in the
continuum (BIC) in the first subband. For increasingly negative strengths, the 
$S$--matrix poles move away again from the real energy axis. The behavior of 
the low-lying pole is particularly worth of attention. As the defect becomes 
more and more attractive, its trajectory bends towards the real axis, which
is reached for $\tilde V_\circ \simeq - 39.6$, at $\tilde E \simeq 0.93$.
At this point, the complex conjugate pole coming from the first quadrant 
collides with the resonance pole, giving rise thereby to a double pole below 
the scattering threshold $\tilde E_n =1$. What is happening can be most 
clearly perceived in the complex $k^{(l)}_1$ plane, as exemplified in Fig. 
\ref{fig4}. As the absolute value of the strength $\tilde V_\circ$ increases, 
the pole in the fourth quadrant of the momentum plane moves downwards 
towards the imaginary axis, the corresponding pole at ${-k^{(l)}_1}^\ast$ 
in the third quadrant doing the same; the two poles collide at 
${\rm Im} k^{(l)}_1 \simeq - 0.83$, when $\tilde V_\circ \simeq -39.6$.
Decreasing the strength further, one has two anti--bound--state poles,
one moving downwards along the negative imaginary axis, the other
moving upwards, until, for $\tilde V_\circ \simeq -46.5$ it crosses the real 
axis, and passes into the upper half of the complex $k$--plane as a bound 
state pole. Had we looked at the corresponding poles in the energy plane, 
we would have seen two poles moving on the real axis on the unphysical 
$(-+++)$ sheet, one of the poles reaching the first scattering threshold
$\tilde E_1$, and going back as a bound--state pole on the physical
$(++++)$ sheet. This situation is strongly reminiscent of what happens in 
standard potential scattering theory, and has been found also in idealized,
one--dimensional models of stubbed waveguides \cite{clp06}. 

We have studied how the bound-state energy changes when the strength of 
the defect or the stub's height $\tilde c$ varies. The possibility of 
bound--state solutions for an empty stub has been known since several years 
ago \cite{lo99}. For an empty stub, the increasing of $\tilde c$ has a 
binding effect, the binding energy $\tilde E_{BS}$ becoming substantially 
independent upon the stub's size for $\tilde c > 2$. This can be easily 
explained, since the bound-state wave function is mainly concentrated in the 
region where the duct matches the stub, and for large cavities a change in the 
transverse dimension cannot affect the wave function in a substantial way. 
When the interaction is switched on, one observes that the bound state 
appears for increasing values of $\tilde c$ for repulsive defects; for a 
given $\tilde c$, the binding energy is shifted upwards. In correspondence 
to an attractive defect, the $(\tilde E_{BS},\; \tilde c)$ curve is shifted 
downwards, and a bound state is possible even for a smooth wave guide, when 
the potential is able to support a bound state by itself.

\begin{figure}
\centerline{\includegraphics[width=8 truecm,angle=-90]{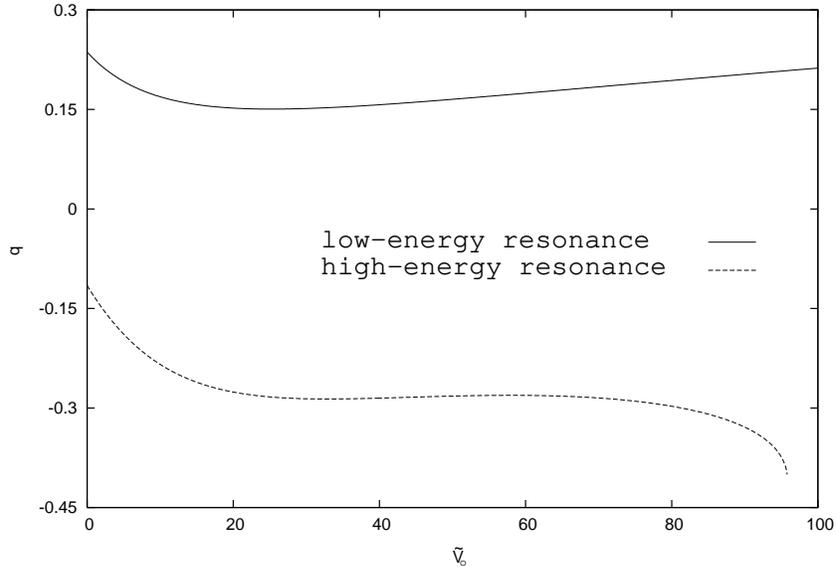}}
\caption{Fano parameter $q$ for the two resonances in the first subband for
$0 \leq \tilde V_\circ \leq 100$ in a waveguide attached to a stub
with $\tilde c = 2$.}
\label{fig5}
\end{figure}

   The asymmetric line shapes one observes in the conductance of quantum
waveguides are generally parametrized in terms of the Fano function
\begin{equation}
T_F = T_0\frac{\left(\epsilon+q \right)^2}{1+\epsilon^2}~~,
\nonumber
\end{equation}
where $T_0$ is the amplitude of the Fano resonance, $\epsilon \equiv 
\left(\tilde E - \tilde E^{(R)}\right)/\tilde \Gamma$ represents the
reduced energy, and the pole location $\tilde E^{(p)}$ in the energy 
plane has been written as $\tilde E^{(p)} \equiv \tilde E^{(R)} - i \tilde
\Gamma$. The Fano parameter $q$ is a measure of the ratio between the resonant
and non--resonant transmission amplitudes, and is related to the
asymmetry of the line shape. For strictly one--dimensional systems, $q$ 
can be evaluated starting from the positions $\tilde E^{(o)}$ 
and $\tilde E^{(p)}$ of the $S$--matrix zeros and poles in the energy plane,
{\em i.e}, \cite{psl93} 
\begin{equation}
q = \frac{\tilde E^{(R)}-\tilde E^{(o)}}{\tilde \Gamma}~~.
\label{fanoq}
\end{equation}
Eq. \ref{fanoq} can be used in the present multi--channel situation below
the second scattering threshold, since in the first subband only one
propagation mode really contributes to the conductance. To this end
we have studied how the zeros of the scattering amplitude move as
the strength varies. As expected, as the interaction becomes more and
more repulsive the zeros move towards the upper threshold, whereas they
are displaced towards lower energies for more and more negative strengths.
In Fig. \ref{fig5} we plot the resulting $q$ parameter for the two 
resonances observed in the first subband, when the strength varies in the
range $0 \leq \tilde V_\circ \leq 100$ and $\tilde c = 2$.   
The Fano parameter of the low--lying resonance is always positive, whereas
for the second resonance it turns out to be negative. This implies
that the two line shapes have opposite asymmetries for all the considered 
strengths, as confirmed by the calculated conductance. In both cases the 
absolute value of the asymmetry parameter remains always different from zero, 
with a rather smooth dependence upon $\tilde V_\circ$ in a rather large range 
of strength values.

\begin{figure}
\centerline{\includegraphics[width=8 truecm,angle=-90]{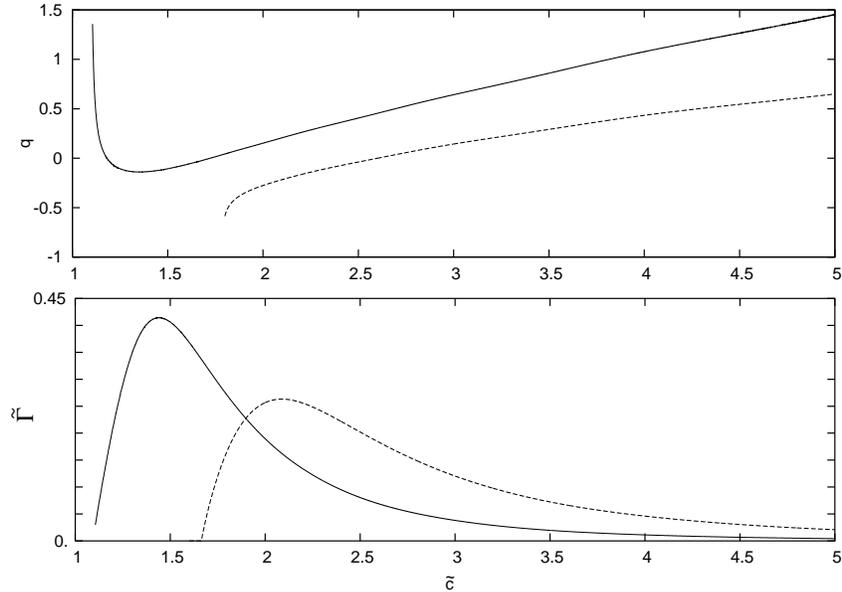}}
\caption{Fano parameter $q$ (upper panel) and width $\tilde \Gamma$ (lower
panel) for the first (solid line) and second (dashed line) resonance in the 
first subband for $1 \leq \tilde c \leq 5$ and $\tilde V_\circ = 20$.}
\label{fig6}
\end{figure}

  We have also studied how the position of the $S$-matrix poles and
zeros depends upon the stub's height $\tilde c$. The ``binding'' 
effect of an increasing $\tilde c$ observed for the bound--state
energies is clearly perceivable in the scattering region also.
As the stub becomes longer, poles and zeros move towards the lower
edge of the first subband, the trajectories of the first and second
pole moving away from the real axis for $\tilde c < 1.5$ and 
$\tilde c < 2.2$, respectively. As $\tilde c$ is increased further,
the two poles come closer and closer to the real axis, giving rise to 
narrower and narrower resonances, until they leave the scattering region.
We looked for the possible presence of BICs with varying $\tilde c$.
No positive-energy states with a vanishing width emerged from our
calculations, both in presence of an impurity and for an empty stub.
BICs for various values of the total transverse width have been
found on the other hand in Ref. \cite{sbr06}, for a quantum billiard
with a transverse hyperbolic profile coupled symmetrically to a quantum
waveguide. There, arguments have been also given in favor of a rather 
widespread presence of BICs in mesoscopic systems. The fact that 
such zero-width states do not appear in the present calculations            
in a rather large range of stub's heights $\tilde c$ $(1 \leq \tilde c
\leq 5)$ can be explained by the geometry we have chosen. Our stub
represents a rather narrow resonator, with $\tilde l_s = 1$; the
quantum billiard considered in Ref. \cite{sbr06}, on the other
hand, has $\tilde l_s = 4$ and $\tilde c \sim 4\div 5$. The
possible presence and location of BICs is strictly related to the
cavity's eigenenergies, whose values and spacings depend in a
crucial way upon the cavity's dimensions. As a matter of fact,
we verified that for an empty stub symmetrically coupled to the leads,
and of dimensions comparable to the quantum billiard of Ref. \cite{sbr06}
there is a BIC at $\tilde E \simeq 1.420$ for $\tilde c = 4.6$, in
remarkable agreement with \cite{sbr06}, given the different shape of
the confining potential. We checked the effect of a host impurity on the 
presence and energy of BICs. To this end, we introduced a double--Gaussian
defect centered in the cavity, having $\tilde l_d = \tilde w =1$. We found 
that BICs are a rather ``robust'' feature of the system, since for 
$\tilde V_\circ =20$ the zero-width state survives the switching on the 
interaction. When the strength is increased up to $\tilde V_\circ = 100$, one 
has a BIC at a slightly higher energy $(\tilde E_{BIC} \simeq 1.465)$, provided
that the stub's height is adjusted to $\tilde c = 4.55$, which means a
variation of the order of $1\%$. Bound states in the continuum may
appear also in more complicated systems. Recently, they have been found
for two open quantum dots, coupled through a connecting bridge \cite{onk06}.
In this case also we verified that when the strength of a host interaction
is increased from $0$ up to $100$, the BIC survives, being slightly
displaced upwards in energy, at price of a small adjustment in the neck's
length $\tilde l_b$. More precisely, one has $\tilde E_{BIC} \simeq 3.30$,
$\tilde l_b = 3.59$ for $\tilde V_\circ = 0$, and $\tilde E_{BIC} \simeq
3.36$, $\tilde l_b = 3.73$ for $\tilde V_\circ = 100$. Both for the
single and the coupled quantum dots the underlying mechanism, as viewed
in the complex energy plane, is the same; when the strength of the
interaction changes, both the pole and the transmission zeros move
away from their original positions and are shifted upwards in energy; since 
they move at different speeds, however, if they coincided for a given value 
of the strength, they do not coincide any longer for the new strength. Their
relative position can be adjusted to zero by a change of some characteristic  
length of the system, so as to have a BIC at a somewhat higher energy.

In analogy with Fig. \ref{fig5}, in Fig. \ref{fig6} we plot the Fano parameter 
$q$ (upper panel) and the width $\tilde \Gamma$ (lower panel) for the 
considered poles as functions of $\tilde c$. Note that the $q$ parameter for 
the higher--energy resonance is not given for $\tilde c < 1.6$; as a matter of 
fact, for stubs of smaller width, this zero-pole pair cannot develop any 
longer in the considered energy region. For stubs having $\tilde c < 1.6$ one 
actually finds that the considered $S$--matrix pole in the $(-+++)$ sheet is 
located above the scattering threshold $\tilde E_2$, thereby becoming a 
shadow pole. The asymmetry parameters turns out to be rather 
sensitive functions of $\tilde c$, and can change sign as $\tilde c$ varies in 
between $1$ and $5$. As a consequence, one expects a rather marked variation 
in the dependence of conductance upon the energy for stubs of different width,
an expectation which is confirmed by our calculations. The origin of 
Fano lineshape reversal has been studied in a model coupled-channel 
calculation in Ref. \cite{kms07}. There, the sign change of $q$ for the
resonances appearing in the electron transmission spectrum of a quantum dot
has been attributed to the coupling among shape and Feshbach resonances 
originating from open and closed channels; this effect turned out to be very 
sensitive to slight modifications of the quantum-dot geometry. A similar 
mechanism is presumably at work here, since as $\tilde c$ increases, the
poles move in the energy plane and new propagating modes open up in the
cavity region; as a consequence. with varying $\tilde c$, the $q$ parameter 
of {\em each} resonance changes sign, and the Fano profiles one observes in 
the first subband may or may not have the same asymmetry, depending upon the 
stub's height one is considering.  

  Finally, we observe that the analysis of the pole trajectories can be 
extended to the $(--++)$ sheet, in correspondence to the second subband. 
Similarly to what has been found below the second threshold, one finds 
a strong repulsive effect with decreasing  $\tilde c$. Poles, determining a 
resonance behavior for a given $\tilde c$, move up in energy until they cross 
the third threshold and become shadow poles; at the same time new poles may 
come into play, moving from lower energies.
     
\section{\label{conc}Conclusions}

  We have studied the poles in the complex energy plane of the multi--channel 
$S$--matrix for a waveguide attached to a resonant cavity. We have allowed
for the presence of a defect in the waveguide. Our main aim was to to get a 
quantitative insight into the dependence of the resonance parameters upon the
defect strength and stub's height. We found that a decrease of the stub's size
as an increase of the positive strength  has a repulsive effect 
on the poles, which move in both cases to higher energies in the $E$--plane.
In particular, as the stub gets shorter, the $S$--matrix poles pass
the upper edge of the considered subband, becoming thereby shadow poles; 
at the same time, poles at lower energies come up, cross the lower threshold 
of the subband, and are therefore able to influence the resonant behavior of 
the conductance. For attractive defects, poles in the first and fourth 
quadrant move towards the real axis as the strength is made more negative; 
with reference to the lowest energy region, there are indications that 
bound states embedded in the continuum occur for critical values of the
strength. As the defect becomes more and more attractive, resonance and 
anti-resonance poles collide on the real axis, giving rise to anti-bound state 
poles moving on opposite directions, until a bound state pole appears on the 
physical sheet, in close analogy with what happens in one--dimensional models 
of stubbed waveguides \cite{clp06}. For repulsive defects, zero-width
scattering states appear for both single and coupled quantum dots, provided
that the resonating cavity has the proper geometry, and represent a 
``robust'' phenomenon with respect to the presence of an embedded interaction. 

 For the lowest subband, where only one open channel contributes to the
conductance, one can relate the parameters appearing in the Fano
function for the line shape to the poles and zeros of the scattering
amplitude in the energy plane. The asymmetry parameter $q$ turns out to be a 
rather smooth function of the strength, whereas it depends more sensitively 
upon the stub's size, and may even change sign with varying $\tilde c$. 
As a consequence, one may or may not observe a reversal of the Fano line 
shape for the resonances in the first subband, depending upon the
value of the stub's height.  

  Some issues deserve certainly further investigations. The role of 
shadow to dominant pole transitions in explaining the changes of
the total conductance near thresholds as $\tilde c$ varies has been    
discussed elsewhere \cite{cl07}. There is evidence of strong resonance 
overlapping in higher subbands. A quantitative study of resonance coupling 
effects, and their dependence upon the dynamical and geometrical parameters of 
the quantum device can be performed resorting to modern effective--Hamiltonian
techniques \cite{op03}, in analogy to what has been done for microwave
billiards \cite{sto02}. These issues are currently under consideration.

\end{document}